\documentclass[twocolumn,superscriptaddress,amsmath,amssymb,aps,prl]{revtex4}
\usepackage{amsmath,amsthm,amssymb}
\usepackage{latexsym}
\usepackage{amscd,graphicx}
\usepackage[titletoc]{appendix}
\usepackage{epsfig}
\usepackage{epstopdf}
\usepackage[normalem]{ulem}
\usepackage[dvipsnames]{xcolor}
\usepackage{booktabs}
\usepackage[colorlinks=true,linkcolor=red,citecolor=blue]{hyperref}

\newcommand{\abs}[1]{| #1 |}

\begin{document}
	
\title{Hybrid Optomechanical Cooling with Kerr Magnons and Squeezed Vacuum}

\author{Xiao-Hong Fan}
\affiliation{Department of Physics, Wenzhou University, Zhejiang 325035, China}

\author{Qin-Geng Chen}
\affiliation{Department of Physics, Wenzhou University, Zhejiang 325035, China}

\author{Jiaojiao Chen}
\altaffiliation{jjchenphys@hotmail.com}
\affiliation{Department of Physics, Wenzhou University, Zhejiang 325035, China}

\author{Wei Xiong}
\altaffiliation{xiongweiphys@wzu.edu.cn}
\affiliation{Department of Physics, Wenzhou University, Zhejiang 325035, China}
\affiliation{International Quantum Academy, Shenzhen, 518048, China}
	

	\date{\today }

	\begin{abstract}
		Ground-state cooling is essential for accessing the quantum regime and enabling quantum control of macroscopic systems. However, achieving optomechanical cooling in the unresolved-sideband regime ($\omega_b < \kappa$) remains challenging. In this Letter, we propose an efficient cooling strategy based on a hybrid optomechanical system incorporating a yttrium iron garnet (YIG) sphere embedded in an optomechanical cavity. Under strong cavity driving, the Kerr nonlinearity of the magnons hosted in the YIG sphere gives rise to a two-magnon process. Adiabatic elimination of the magnons yields an effective two-photon process in the cavity, which destructively interferes with all dissipative channels, surpassing the quantum backaction limit and enabling \textit{complete suppression} of heating under optimal conditions, even in the deeply unresolved sideband regime (DUSR: $\omega_b \ll \kappa$). Moreover, injecting squeezed vacuum noise into the cavity not only preserves these advantages but also delivers additional enhancements, including higher net cooling rates, reduced optomechanical coupling requirements, and improved noise robustness. Comparative analysis shows that our approach outperforms existing schemes without Kerr magnons, underscoring the potential of integrating nonlinear magnonics with optomechanics for quantum control of macroscopic mechanical systems.
	\end{abstract}

	\maketitle
	
{\it Introduction.}---Quantum control of macroscopic objects remains a long-standing pursuit in quantum science~\cite{Bocko1996,Knobel2003}, where suppressing thermal motion via cooling is essential to reveal quantum phenomena. Diverse cooling techniques have been developed over decades, particularly in atomic physics~\cite{Phillips1998,Wineland2013}. Recently, mechanical resonators operating in the quantum regime~\cite{Kippenberg2008,Aspelmeyer2014} have emerged as promising platforms for quantum information~\cite{OConnell2010,Verhagen2012,Wang2012,Dong2012}, quantum-limited measurements~\cite{LaHaye2004,Teufel2009,McClelland2011,Krause2012,Purdy2013,Li2018}, and fundamental tests of quantum mechanics~\cite{RomeroIsart2011,Pepper2012,Blencowe2013,Sekatski2014}. In cavity optomechanics, ground-state cooling is achievable through red-sideband driving~\cite{WilsonRae2007,Marquardt2007} and has been experimentally demonstrated~\cite{Gigan2006,Arcizet2006,Rocheleau2010,Teufel2011,Chan2011,Peterson2016}. However, quantum backaction from the photon-phonon parametric interaction imposes a fundamental limit on the minimum phonon number, scaling inversely with the sideband resolution $\omega_b/\kappa$ (where $\omega_b$ is the mechanical frequency and $\kappa$ is the cavity decay rate). Overcoming this limit is particularly challenging in the unresolved sideband regime (USR: $\omega_b<\kappa$), despite proposed approaches including pulsed driving~\cite{Machnes2012,Wang2011}, dissipative coupling~\cite{Elste2009,Li2009b,Xuereb2011}, squeezed driving~\cite{Clark2017}, feedback control~\cite{Rossi2017,Zippilli2018}, and hybrid schemes~\cite{Liu2015,Gu2013,Guo2014,Ojanen2014,Chen2015,Genes2009}. Moreover, a method to fully eliminate backaction heating remains elusive, as cavity dissipation cannot be completely suppressed with current techniques~\cite{Weiss2013}.

With the rise of quantum materials, magnons (i.e., collective spin excitations) in yttrium iron garnet (YIG) spheres, known for their high spin density and tunable properties, have been paid consideration attention in quantum information science~\cite{Rameshti2022,Yuan2022}. In particular, Kerr magnons (magnons with Kerr nonlinearity), originating from magnetocrystalline anisotropy~\cite{Zhang2019} and experimentally realized in YIG spheres~\cite{Wang2016,Wang2018}, enable a wealth of nonlinear physics, including bistability~\cite{Wang2016,Wang2018,Shen2021,Shen2022}, nonreciprocal entanglement~\cite{Chen2023,Chen2024}, strong spin coupling~\cite{Xiong2022,Ji2023}, blockade~\cite{Fan2024,Hou2024}, quantum phase transitions~\cite{Zhang2021,Liu2023}, steady-state squeezing~\cite{Qi2025}, and other nonlinear effects~\cite{Zheng2023}. Leveraging these capabilities, we propose an efficient cooling scheme based on a micron-sized YIG sphere embedded in an optomechanical cavity. Under strong cavity driving, the magnon Kerr effect gives rise to a two-magnon process which, after adiabatic elimination of the magnon mode, manifests as an effective two-photon process in the cavity. When the cavity couples to a thermal bath~\cite{Walls2008}, this induced two-photon interaction can destructively interfere with cavity dissipation channels, thereby overcoming the sideband-resolution constraint imposed by quantum backaction and enabling ground-state cooling deep in the unresolved-sideband regime. At optimal cooling conditions, backaction heating is fully suppressed, yielding a significantly reduced minimum phonon occupancy. Moreover, injecting squeezed vacuum noise into the cavity not only preserves these advantages but also offers further enhancements, including higher net cooling rates, lower optomechanical coupling requirements, and greater robustness against noise. Fine-tuning the two-photon interaction strength further improves cooling efficiency, surpassing both standard sideband cooling scheme (SB) ~\cite{WilsonRae2007} and squeezed cooling scheme (SS)~\cite{Clark2017,Asjad2016}. This work establishes a new pathway for quantum control of macroscopic mechanical systems via the hybridization of Kerr magnonics and optomechanics. To facilitate reading, we list the several abbreviations used in this article in Table~\ref{table}.

\begin{table}
	\centering
	\caption{Abbreviations and Definitions}
	\begin{tabular}{cc}
		\hline
		\textbf{Abbreviation} & \textbf{Definition} \\
		\hline
		YIG  & Yttrium Iron Garnet \\
		USR  & Unresolved Sideband Regime \\
		DUSR & Deeply Unresolved Sideband Regime \\
		SB   & Sideband Cooling Scheme \\
		KS   & Kerr-Magnon Cooling Scheme \\
		SS   & Squeezed Cooling Scheme \\
		HS   & Hybrid Cooling Scheme \\
		\hline
	\end{tabular}\label{table}
\end{table}

\begin{figure}
\includegraphics[scale=0.5]{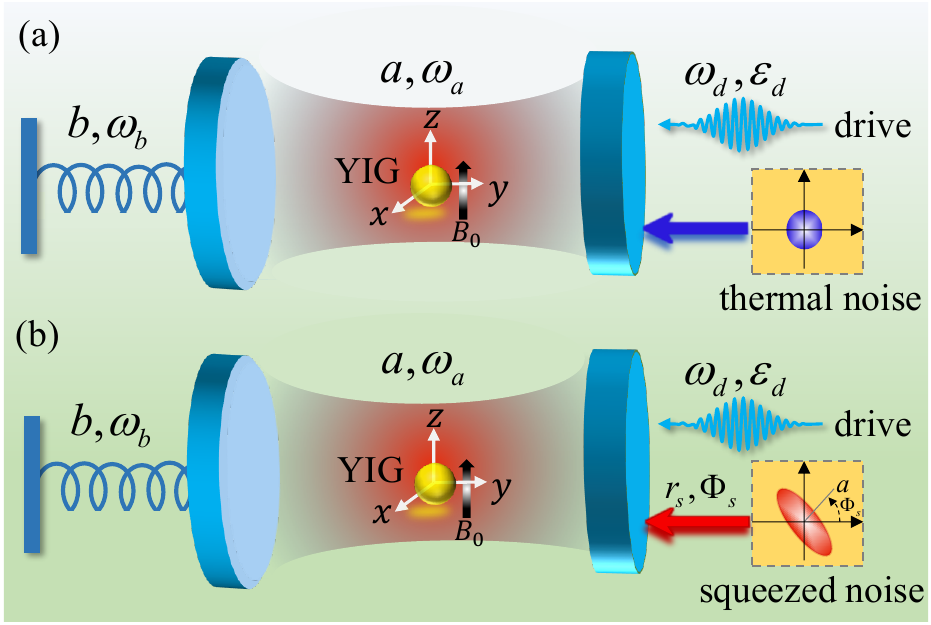}
\caption{Schematic diagram of the proposed hybrid quantum system.
	A yttrium iron garnet (YIG) sphere supporting Kerr magnons is placed inside a driven optomechanical cavity and subjected to a static magnetic field \( B_0 \) (black arrow) aligned along the crystallographic \( z \)-axis. 
	(a) The cavity is coupled to a thermal reservoir (blue circle) with trivial (white) noise correlations. 
	(b) The cavity is instead coupled to a broadband squeezed vacuum bath (red ellipse) with nontrivial correlations, characterized by squeezing parameter \( r_s \) and phase \( \Phi_s \). 
	Here, \( \omega_a \), \( \omega_b \), and \( \omega_d \) denote the frequencies of the cavity photons, mechanical mode, and drive, respectively, and \( \varepsilon_d \) is the amplitude of the drive field.}
\label{fig1}
\end{figure}

{\it The model and Hamiltonian.}---We consider a hybrid quantum system comprising a micron-sized YIG sphere embedded in a driven optomechanical cavity with frequency $\omega_a$~[see Fig.~\ref{fig1}(a)]. In a rotating frame at the drive frequency $\omega_d$, the total Hamiltonian is given by (setting $\hbar = 1$) $H = H_{\rm OM} + H_{K} + H_{I} + H_D$, where $H_{\rm OM} = \delta_a a^\dag a + \omega_b b^\dag b + g_0 a^\dag a(b + b^\dag)$ is the Hamiltonian of the simplest optomechanical system, with $\delta_a=\omega_a-\omega_d$ the cavity-drive detuning, $\omega_b$ the mechanical frequency, $a$ ($b$) and $a^\dag$ ($b^\dag$) the photon (phonon) annihilation and creation operators, and $g_0$ the single-photon optomechanical coupling strength. The second term, $H_K = \delta_m m^\dag m - (K/2) m^\dag m^\dag m m$, is the Hamiltonian of Kerr magnons, where $\delta_m = \omega_m - \omega_d$ with $\omega_m$ the magnon frequency is the magnon-drive detuning, $m$ ($m^\dag$) is the magnon annihilation (creation) operator, and $K$ characterizes the strength of the Kerr nonlinearity, arising from the magnetocrystalline anisotropy, which can be well controlled by the static magnetic field $B_0$~\cite{Wang2016,Wang2018}. The third term, $H_I = J(a^\dag m + a m^\dag)$, captures the photon-magnon interaction, with coupling strength $J$. The last term, $H_D = \varepsilon_d(a^\dag + a)$, models a coherent field, with $\varepsilon_d$ being the drive amplitude. For a strong drive, the system can be linearized around its steady state, leading to frequency shifts on both the cavity mode and the Kittel mode (see the Supplementary Material~\cite{SM} for details), i.e., $\Delta_a = \delta_a + g_0(b_s + b_s^*)$ and $\Delta_m = \delta_m - 2K|m_s|^2$, with $b_s$ and $m_s$ steady-state values. Additionally, the optomechanical coupling is correspondingly enhanced to $G = g_0 a_s$ and the Kerr nonlinearity is transformed into a two-magnon process ($\propto m^\dag m^\dag+mm$) with tunable coefficient $K_m = K m_s^2$~\cite{SM}. By adiabatically eliminating the degrees of freedom of the rapidly decohering Kerr magnons, the effective optomechanical Hamiltonian can be obtained~\cite{SM},
\begin{align}
H_{\rm eff} = \Delta a^\dag a + \omega_b b^\dag b + G(a + a^\dag)(b + b^\dag) + H_{{\bf A}^2}, \label{eq2}
\end{align}
where $\Delta = \Delta_a - \eta \Delta_m$ is the effective frequency of the cavity with the effective decay rate $\kappa = \kappa_a + \eta \kappa_m$,  with $\eta = J^2 / (\Delta_m^2 + \kappa_m^2/4 - |K_m|^2)$. The last term, $H_{{\bf A}^2} = \xi^* a^2 + \xi a^{\dag 2}$ with $\xi = -\eta K_m / 2$, represents the two-photon process induced by Kerr magnons.

\begin{figure}
 	\includegraphics[scale=0.18]{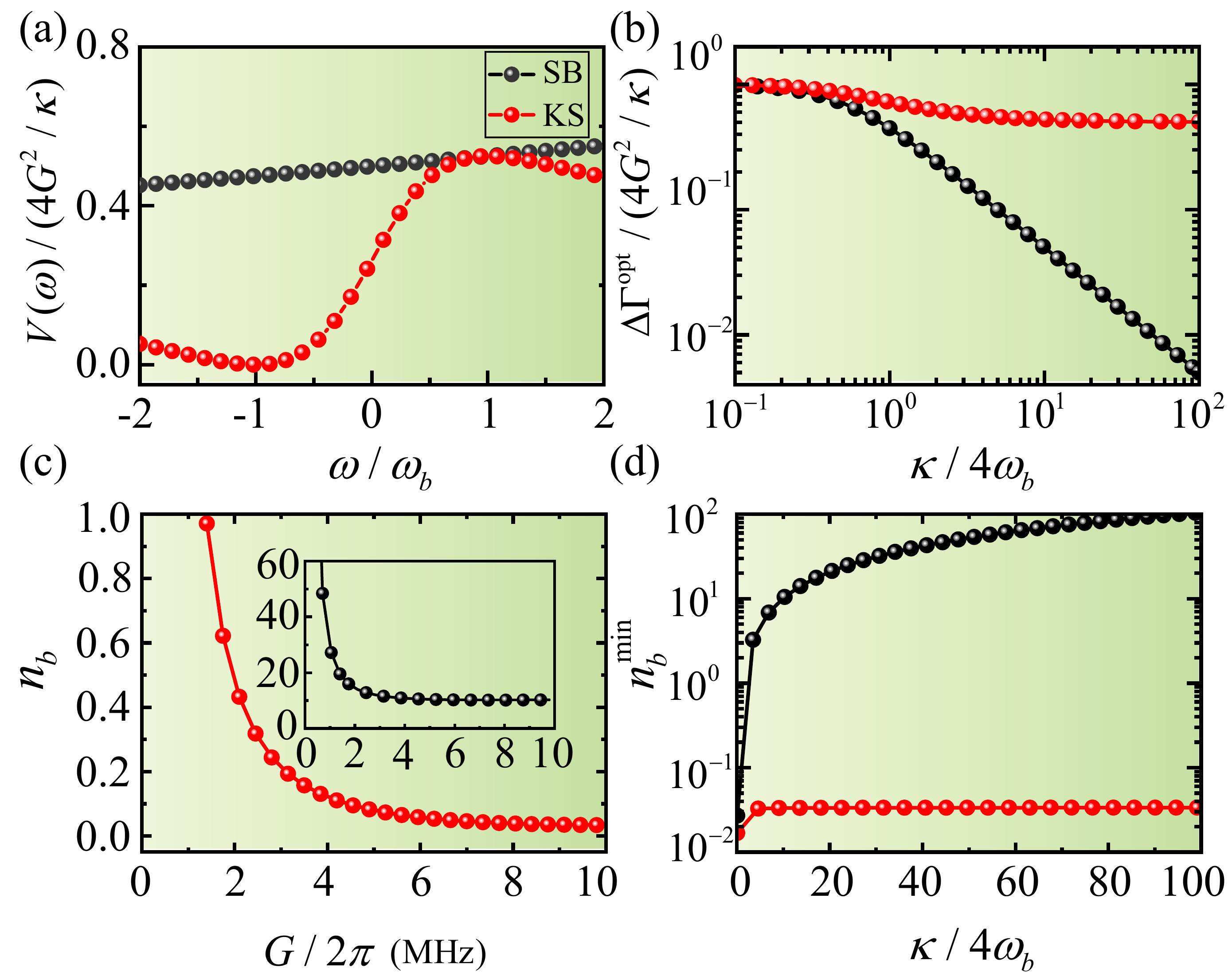}
 	\caption{(a) The normalized radiation-pressure spectra of SB (black) and KS (red) as functions of the normalized frequency $\omega/\omega_b$, with $\kappa/4\omega_b=10$. 
 		(b) The normalized optimal net cooling rates of SB (black) and KS (red) as functions of the normalized cavity decay $\kappa/4\omega_b$. 
 		(c) The final phonon occupancy $n_b$ of SB (inset) and KS (red) as functions of  the coupling strength $G/2\pi$, with $\kappa/4\omega_b=10$. 
 		(d) The minimum phonon number $n_b^{\rm min}$ of SB (black) and KS (red) as functions of the normalized cavity decay $\kappa/4\omega_b$. Other parameters are set as $\omega_b/2\pi = 10~\mathrm{MHz}$, $\gamma_b/2\pi = 10~\mathrm{Hz}$, and $\Delta = \sqrt{\kappa^2/4 + \omega_b^2}$.}\label{fig2}
 \end{figure}

{\it Optomechanical cooling with Kerr magnons.}---When Kerr magnons are introduced (i.e., $\xi \neq 0$), we term the scheme as Kerr-magnon cooling scheme (KS). In the weak-coupling regime ($G \ll \kappa$), the radiation-pressure spectrum is given by~\cite{SM} $V_{\rm KS}(\omega) = V_{\rm SB}(\omega) \abs{1 - 2i \xi \chi(-\omega)}^2 / \abs{1 - 4|\xi|^2 \chi(\omega) \chi^*(-\omega)}^2$, where $V_{\rm SB}(\omega) = G^2 \kappa x_{\rm ZPF}^{-2} |\chi(\omega)|^2$ is the spectrum of SB, $\chi(\omega) = [\kappa/2 - i(\omega - \Delta)]^{-1}$ is the cavity susceptibility, and $x_{\rm ZPF} = \sqrt{\hbar / (2 m_{\rm eff} \omega_b)}$ is the zero-point fluctuation amplitude of the mechanical oscillator with effective mass $m_{\rm eff}$. According to Fermi's golden rule~\cite{Clerk2010}, the cooling and heating rates are defined as $\Gamma_{\rm KS}^- = V_{\rm KS}(\omega_b)$ and $\Gamma_{\rm KS}^+ = V_{\rm KS}(-\omega_b)$, respectively. The net cooling rate is then given by $\Delta \Gamma_{\rm KS} = \Gamma_{\rm KS}^- - \Gamma_{\rm KS}^+ = [|1 - 2i\xi \chi(-\omega_b)|^2 \Gamma_{\rm SB}^- - |1 - 2i\xi \chi(\omega_b)|^2 \Gamma_{\rm SB}^+] / |1 - 4|\xi|^2 \chi(-\omega_b) \chi^*(\omega_b)|^2$, where $\Gamma_{\rm SB}^- = V_{\rm SB}(\omega_b) = G^2 \kappa / [\kappa^2/4 + (\omega_b - \Delta)^2]$ and $\Gamma_{\rm SB}^+ = V_{\rm SB}(-\omega_b) = G^2 \kappa / [\kappa^2/4 + (\omega_b + \Delta)^2]$ are the cooling and heating rates in SB. Figure~\ref{fig2}(a) shows that the radiation-pressure spectrum becomes highly asymmetric in KS, indicating a substantial net cooling effect. In contrast, the spectrum in SB is nearly symmetric, resulting in a vanishing net cooling rate. This asymmetry arises from quantum interference between the two-photon process and the cavity dissipation channels. Notably, at $\omega = -\omega_b$, destructive interference completely suppresses the heating process, i.e., $\Gamma_{\rm KS}^+ = V_{\rm KS}(-\omega_b) = 0$, directly giving rise to an optimal cooling condition
\begin{equation}
	\xi_{\rm KS} = \frac{\Delta - \omega_b}{2} - i \frac{\kappa}{4}. \label{eq6}
\end{equation} 
When $\xi=\xi_{\rm KS}$, the net cooling rate is optimized as $\Delta\Gamma_{\rm KS}^{\rm opt} = \Gamma_{\rm SB}^-$. Due to unvanishing heat rate $\Gamma_{\rm SB}^+$, $\Delta\Gamma_{\rm KS}^{\rm opt}$ is naturally larger than the net cooling rate $\Delta\Gamma_{\rm SB}$ in SB ($\xi=0$). Notably, $\Delta\Gamma_{\rm KS}^{\rm opt}$ is approximately enhanced by two orders of magnitude over $\Delta\Gamma_{\rm SB}$ in DUSR ($\kappa / 4\omega_b = 100$), i.e., $\Delta\Gamma_{\rm KS}^{\rm opt} / \Delta\Gamma_{\rm SB} \approx 10^2$, as shown by Fig.~\ref{fig2}(b).
		
Following the quantum noise approach, the final phonon number can be expressed as $n_b = n_c + n_q$, where $n_c = n_{\rm th} \gamma_b / \Delta\Gamma_{\rm KS}$, with mechanical damping rate $\gamma_b$ and the thermal phonon occupation $n_{\rm th}$, denotes the classical cooling limit, and $n_q = \Gamma_{\rm KS}^+ / \Delta\Gamma_{\rm KS}$ represents the quantum limit. In SB, $n_q=\kappa / 4\omega_b$ at the optimal detuning $\Delta = \sqrt{\kappa^2/4 + \omega_b^2}$. This indicates that SB is fundamentally constrained by the sideband resolution. Therefore, SB cannot be achieved in USR [see inset in Fig.~\ref{fig2}(c)]. However, when Kerr magnons are included, the quantum limit $n_q=0$ at $\xi=\xi_{\rm KS}$ because of $\Gamma_{\rm KS}^+=0$. This suggests that KS is no longer constrained by the sideband resolution, enabling ground-state cooling in USR, as illustrated in Fig.~\ref{fig2}(c). Notably, the final phonon number can be further significantly suppressed to $n_b \ll 1$ by slightly increasing the optomechanical coupling strength $G$, as also demonstrated in Fig.~\ref{fig2}(c). Figure~\ref{fig2}(d) further reveals that the final phonon number $n_b^{\rm min}$ at $\xi=\xi_{\rm KS}$, i.e., the minimum phonon number $n_b^{\rm min}$, is remarkably robust against cavity dissipation when Kerr magnons are included, maintaining efficient cooling ability even in DUSR. By contrast, $n_b^{\rm min}$ in SB is extremely sensitive to the cavity decay rate, rapidly leading to $n_b^{\rm min}>1$ in USR.
 	   
\begin{figure}
	\includegraphics[scale=0.18]{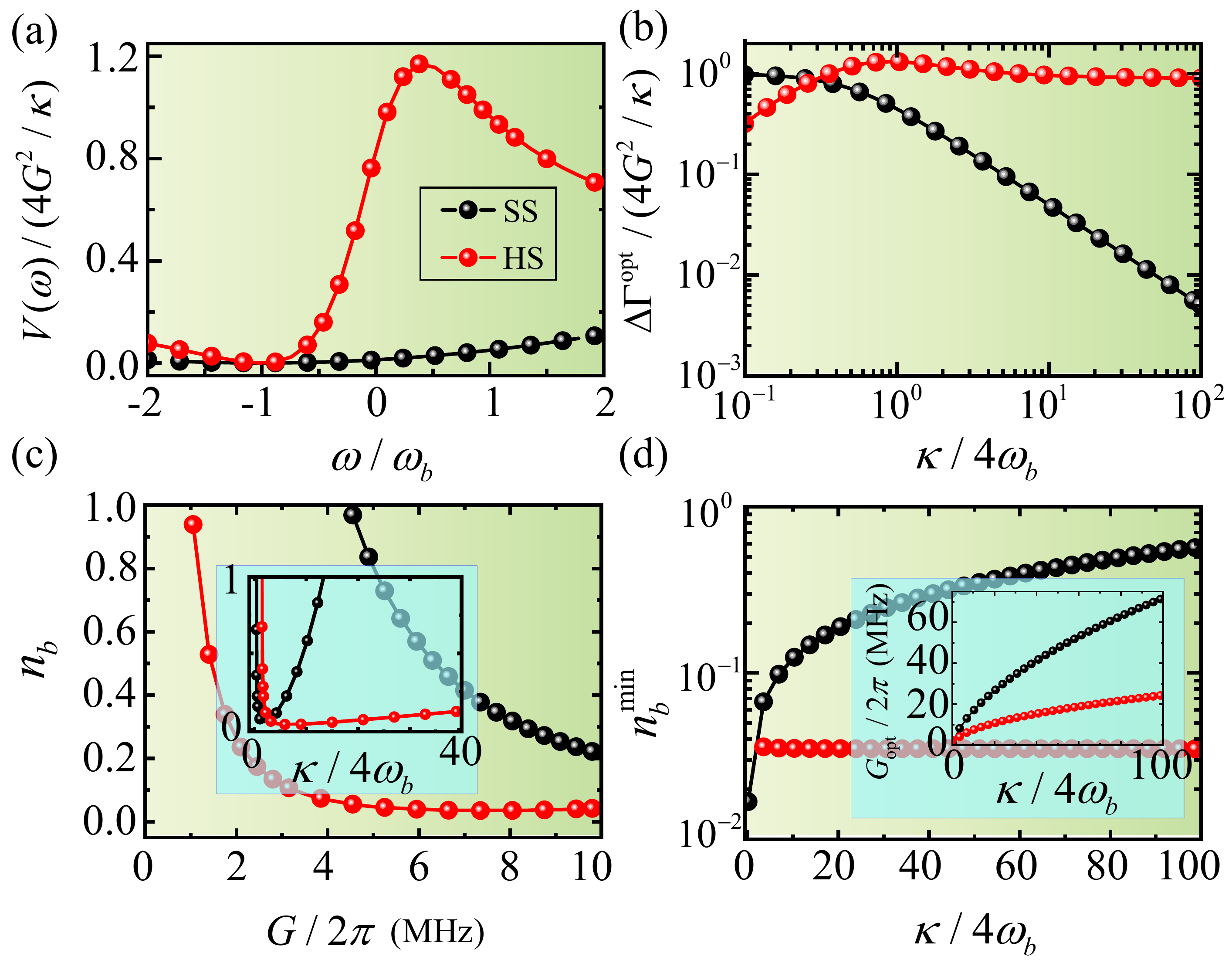}
	\caption{(a) The normalized radiation-pressure spectra of SS (black) and HS (red) as functions of the normalized frequency $\omega/\omega_b$, with $\kappa/4\omega_b=10$. 
		(b) The normalized optimal net cooling rates of SS (black) and HS (red) as functions of the normalized cavity decay $\kappa/4\omega_b$. 
		(c) The final phonon occupancy $n_b$ of SS (black) and HS (red) as functions of  the coupling strength $G/2\pi$, with $\kappa/4\omega_b=10$. Inset: The final phonon occupancy $n_b$ of SS (black) and HS (red) as functions of  the normalized cavity decay $\kappa/4\omega_b$, with $G/2\pi=6$ MHz.  
		(d) The minimum phonon number $n_b^{\rm min}$ of SS (black) and HS (red) as functions of the normalized cavity decay $\kappa/4\omega_b$. Inset: The optimal optomechanical coupling strength as functions of the normalized cavity decay $\kappa/4\omega_b$, where $\abs{\xi_{\rm HS}}=13.92\omega_b$. Other parameters are the same as in  Fig.~\ref{fig2}.}\label{fig3}
\end{figure}

{\it Hybrid cooling with both Kerr magnons and squeezed bath.}---We below study hybrid cooling scheme (HS) by injecting a squeezed vacuum noise~\cite{Ast2013,Serikawa2016,Clark2016,Murch2013} into the optomechanical cavity including Kerr magnons [Fig.~\ref{fig1}(b)]. Unlike thermal noise in Fig.~\ref{fig1}(a), the squeezed noise exhibits nontrivial correlations~\cite{Breuer2002}: $\langle a_{\rm in}^\dagger(\omega)a_{\rm in}(\omega^\prime)\rangle=\mathcal{N}_s\delta(\omega+\omega^\prime)$, $\langle a_{\rm in}(\omega)a_{\rm in}^\dagger(\omega^\prime)\rangle=(\mathcal{N}_s+1)\delta(\omega+\omega^\prime)$, $\langle a_{\rm in}(\omega)a_{\rm in}(\omega^\prime)\rangle=\mathcal{M}_s\delta(\omega+\omega^\prime)$, and $\langle a_{\rm in}^\dagger(\omega)a_{\rm in}^\dagger(\omega^\prime)\rangle=\mathcal{M}_s^*\delta(\omega+\omega^\prime)$, with $\mathcal{N}_s=\sinh^2 r_s$ and $\mathcal{M}_s=e^{-2i\Phi_s}\sinh r_s\cosh r_s$. These correlations modify the radiation-pressure spectrum to
\begin{equation}
   {V_{\rm HS}(\omega)}={V_{\rm KS}(\omega)}\left|\cosh r_s + A_\xi(\omega) e^{-2i\Phi_s} \sinh r_s \right|^2, \label{eq3}
\end{equation}
where $A_\xi(\omega)=A_0(\omega)[1+2i\xi^*\chi^*(\omega)]/[1-2i\xi\chi(-\omega)]$ with $A_0(\omega)=\chi(-\omega)/\chi^*(\omega)$. Without Kerr magnons ($\xi=0$), $A_\xi(\omega)=A_0(\omega)$, $V_{\rm KS}(\omega)=V_{\rm SB}(\omega)$, thus, $V_{\rm HS}(\omega)= V_{\rm SB}(\omega)\left|\cosh r_s + A_0(\omega) e^{-2i\Phi_s} \sinh r_s\right|^2=V_{\rm SS}(\omega)$, which is just the radiation-pressure spectrum ($V_{\rm SS}$) of SS. For comparision, both $V_{\rm SS}$ (black) and $V_{\rm HS}$ (red) are plotted in Fig.~\ref{fig3}(a). Obviously, $V_{\rm HS}(\omega)$ displays a sharp change across the frequency domain in USR ($\kappa/4\omega_b = 10$), while $V_{\rm SS}$ nearly keeps unchanged. The abrupt variation in $V_{\rm HS}$ here is caused by the quantum interference effect, arising from the two-photon process and cavity dissipation channels. This finding aligns well with the features observed in Fig.~\ref{fig2}(a). As depicted in Fig.~\ref{fig3}(a), the heating rate $\Gamma_{\rm HS}^+\propto V_{\rm HS}(\omega)$ vanishes at $\omega=-\omega_b$, pointing to destructive interference at this frequency. This vanishing of the heat rate directly results in
\begin{equation}
 \tanh r_s\, e^{-2i\, \phi_s} +A_{\xi_{\rm HS}}^*(\omega_b)=0, \label{eq5}
\end{equation}
which is the optimal cooling condition of HS. For SS, the optimal cooling condition is obtained by setting $\xi = 0$ in Eq.~(\ref{eq5}), yielding $\tanh r_s e^{-2i\phi_s} = -A_0^*(\omega_b)$. When the respective optimal conditions for two schemes are satisfied, heating processes are {\it completely suppressed} in both cases. As a result, the net cooling rates reach their optimal values: $\Delta\Gamma_{\rm SS}^{\rm opt} = \Delta\Gamma_{\rm SB}$ for SS and $\Delta\Gamma_{\rm HS}^{\rm opt} = \Gamma_{\rm KS}^- \cosh^2 r_s$ for HS. This demonstrates that HS achieves an {\it exponentially enhanced} cooling rate compared to $\Gamma_{\rm KS}^-$ in KS, while SS can only reach the limit of SB.

Unlike SS, $\Delta\Gamma_{\rm HS}^{\rm opt}$ exhibits only slight fluctuations as the cavity drive shifts from USR to DUSR. Conversely, $\Delta\Gamma_{\rm SS}^{\rm opt}$ decreases rapidly, as clearly shown in Fig.~\ref{fig3}(b). This difference becomes particularly pronounced in DUSR, highlighting the superior robustness and performance of HS. Figure~\ref{fig3}(c) further demonstrates the superiority of our HS over SS in terms of the required optomechanical coupling strength $G$ within USR ($\kappa/4\omega_b = 10$). To achieve comparable cooling performance, HS requires a significantly weaker coupling strength than SS. At a fixed coupling strength of $G/2\pi=6$ MHz [see the inset of Fig.~\ref{fig3}(c)], HS not only achieves a lower final phonon occupation but also maintains efficiency as the system transitions from USR to DUSR. In contrast, SS operates reliably only within USR. These results are consistent with the observations in Fig.~\ref{fig3}(b). Figure~\ref{fig3}(d) further shows that the minimum achievable phonon number in HS is remarkably robust against cavity dissipation, whereas in SS it increases rapidly. The inset in Fig.~\ref{fig3}(d) indicates that the required optimal coupling strength $G_{\rm opt}$ in HS is significantly smaller than that in SS. 
 
All analytical results above are derived under the weak-coupling approximation and verified by exact master-equation simulations, as detailed in the Supplemental Material~\cite{SM}. Excellent agreement is found across regimes, along with a full stability analysis~\cite{Gradshteyn1980} and parameter maps for ground-state cooling. HS remains effective in DUSR and supports higher bath temperatures than other schemes, underscoring its robustness and experimental feasibility.

\begin{figure}
	\includegraphics[scale=0.195]{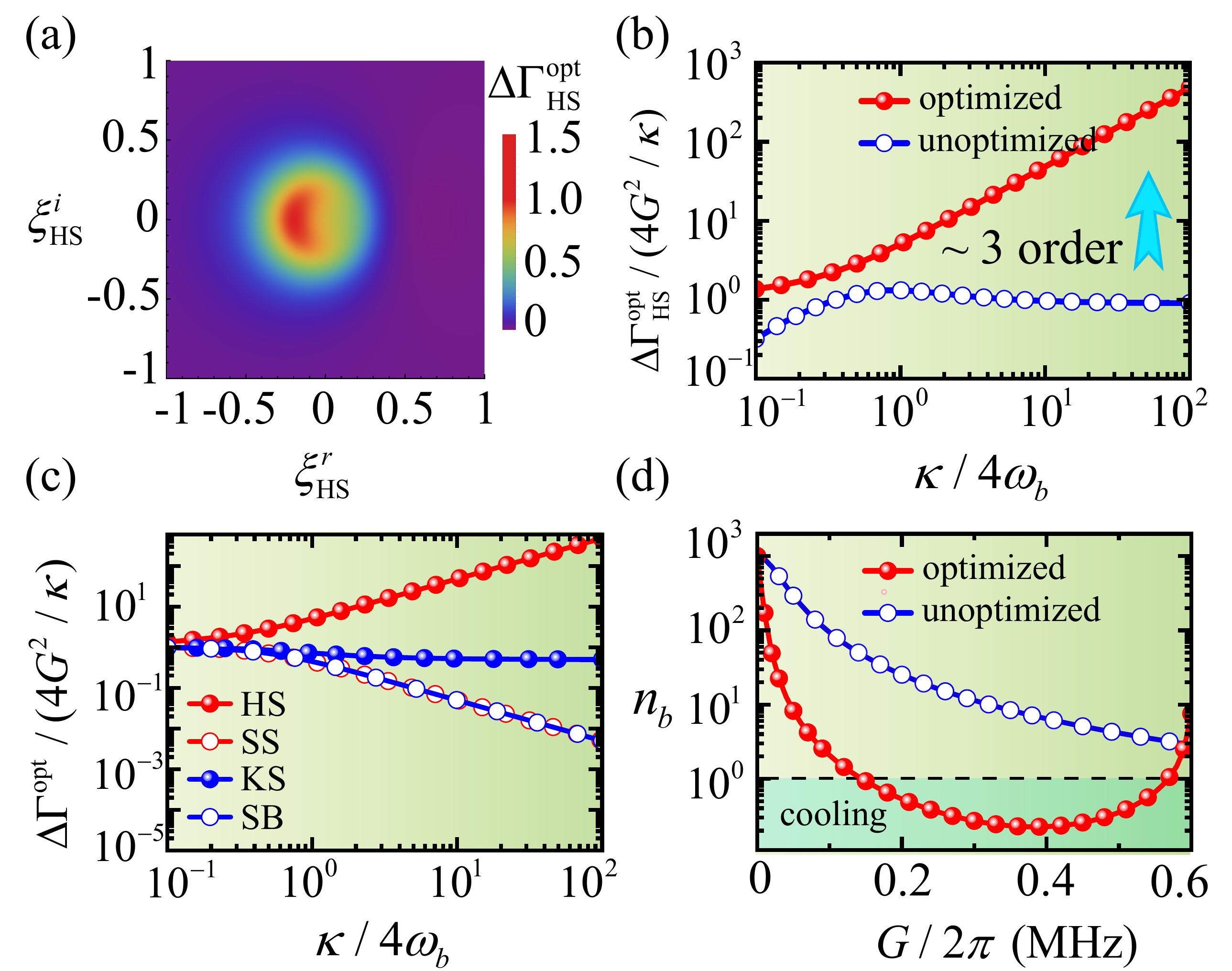}
	\caption{(a) The optimal net cooling rate ${\Delta\Gamma^{\rm opt}_{\rm HS}}$ as functions of the real part ($\xi _{\rm HS}^r$) and the imaginary part ($\xi _{\rm HS}^i$), with $(\kappa /4{\omega _b} = 0.1)$. (b) The unoptimized (blue) and optimized (red) net cooling rate ${\Delta\Gamma^{\rm opt}_{\rm HS}}$ of HS as a function of the normalized cavity decay $\kappa/4\omega_b$. (c) The normalized net cooling rates of HS (red solid circles), SS (red open circles), KS (blue solid circles), and SB (blue open circles), as functions of the normalized cavity decay $\kappa /4{\omega _b}$. (d) The phonon number $n_b$ in HS as a function of the coupling strength $G/2\pi$ for $\kappa /4{\omega_b} = 10$. Blue (red) curves correspond to unoptimized (optimized) values of $\xi_{\rm HS}$. Other parameters are the same as those in Fig.~\ref{fig2}.}\label{fig4}
\end{figure}  
{\it Cooling enhancement via optimization.}---Under the optimal cooling condition of HS given by Eq.~(\ref{eq5}), the optimal net cooling rate is achieved. However, the parameter $\xi$ introduced in Eq.~(\ref{eq5}) remains a free variable, offering further potential for optimization. To exploit this degree of freedom, we express the parameter $\xi_{\rm HS}$ as $\xi_{\rm HS} = \xi_{\rm HS}^r + i \xi_{\rm HS}^i$ and numerically explore the dependence of the net cooling rate on both the real and imaginary parts of $\xi$ for various values of $\kappa / 4\omega_b$ [see Fig.~\ref{fig4}(a)]. Our results demonstrate that the cooling performance can be significantly improved by jointly optimizing $\xi_{\rm HS}^r$ and $\xi_{\rm HS}^i$. For example, when $\kappa / 4\omega_b = 0.1$, the net cooling rate reaches an optimal value of $\Gamma_{\rm HS}^{\rm opt} = 1.36\omega_b$ at $\xi_{\rm HS}^r = -0.196\omega_b$ and $\xi_{\rm HS}^i = 0$. Similar optimizations performed across other parameter regimes yield consistently enhanced cooling rates, as summarized in the Supplementary Material~\cite{SM}. Remarkably, the advantage of this optimization becomes particularly pronounced in DUSR. For instance, at $\kappa / 4\omega_b = 100$, the optimized net cooling rate is enhanced by approximately {\rm three orders} of magnitude compared to the unoptimized case [see arrow in Fig.~\ref{fig4}(b)].
      
To further highlight the superiority of our optimized HS, we compare its performance with SB, KS, and SS. As shown in Fig.~\ref{fig4}(c), the optimized cooling rate in HS significantly outperforms the others, especially as the system transitions from USR to DUSR. Specifically, while the cooling rates in SS and SB decrease, and that in KS remains nearly constant with increasing $\kappa / 4 \omega_b$, the cooling rate in HS continues to grow nearly linearly. In addition to improving the cooling rate, the optimization of $\xi$ drastically reduces the required optomechanical coupling strength to achieve efficient cooling, as shown in Fig.~\ref{fig4}(d). These results underline the critical importance of optimizing the parameter $\xi$ to maximize cooling efficiency, particularly in regimes where cooling in SB become ineffective. The pronounced enhancement underscores the potential of finely tuning the two-photon process coefficient, introduced by Kerr magnons, as a powerful and practical approach to enhance quantum control in hybrid optomechanical platforms. 
 
{\it Possible implementations.}---The proposed protocols can be implemented in a three-dimensional (3D) electromechanical platform~\cite{Liu2025}, consisting of a 3D superconducting cavity coupled to a mechanical resonator via a parallel-plate capacitor. With a typical internal volume of $\sim\text{mm}^3$, such cavities can readily accommodate a millimeter-scale yttrium iron garnet (YIG) sphere, as commonly used in cavity magnonics experiments~\cite{Huebl2013,Zhang2014}. This makes the integration of 3D electromechanics with magnonic systems experimentally accessible using current technologies. A key requirement of our scheme is the introduction of magnon Kerr nonlinearity, which can be realized by applying a static magnetic field along specific crystallographic axes (e.g., $[100]$ or $[110]$) of the YIG sphere~\cite{Wang2016,Wang2018}. High-resolution tunable electromagnets allow precise control of this static field, enabling fine-tuning of the magnonic properties. Our analysis also relies on the adiabatic elimination of the magnon mode, which requires relatively large magnon damping. This condition can be met by engineering additional losses in the YIG sphere, through either rare-earth ion doping~\cite{Huebl2013} or surface deposition of iron particles to enhance scattering~\cite{Zhang2014}. The required squeezed vacuum field for the 3D microwave cavity can be generated using a Josephson parametric amplifier (JPA)~\cite{CastellanosBeltran2008,Clark2016,Clark2017,Murch2013}, a well-established technique in circuit quantum electrodynamics. JPAs enable the generation of broadband, tunable squeezed microwaves, providing the flexibility needed to optimize cooling performance as demonstrated in our work. Together, squeezed fields, Kerr magnons, and 3D optomechanics constitute a robust and accessible hybrid platform. The proposed scheme offers a realistic pathway for exploring quantum phenomena in hybrid optomechanical systems and for developing next-generation hybrid quantum devices.

{\it Conclusion.}---We propose an efficient scheme for ground-state cooling in DUSR using a hybrid optomechanical system containing an embedded yttrium iron garnet (YIG) sphere. The Kerr magnons in the YIG sphere act as an auxiliary subsystem that induces effective two-photon process, enabling destructive interference with cavity dissipation channels. This yields \textit{complete suppression} of phonon heating under optimal conditions, achieving ground-state cooling in this challenging regime. By injecting squeezed vacuum noise into the cavity, we obtain additional advantages, including higher net cooling rates, reduced optomechanical coupling requirements, and improved noise resilience. Comparative analysis demonstrates that our proposal substantially advances previous schemes without Kerr magnons. These results highlight the potential of integrating nonlinear magnons with optomechanics for quantum control of macroscopic mechanical systems.

This work was supported by the Natural Science Foundation of Zhejiang Province (GrantNo. LY24A040004), Zhejiang Province Key R\&D Program of China (Grant No. 2025C01028),  and Shenzhen International Quantum Academy (Grant No. SIQA2024KFKT010).


\end{document}